\documentclass[12pt]{article}
\usepackage{amsfonts}
\newcommand{\vf}{\varphi}
\newcommand{\be}{\begin{equation}}
\newcommand{\ee}{\end{equation}}
\newcommand{\ba}{\begin{eqnarray}}
\newcommand{\ea}{\end{eqnarray}}
\newcommand{\no}{\nonumber\\}
\newcommand{\bi}{\bibitem}
\newcommand{\vect}[1]{\stackrel{\rightarrow}{#1}}
\def\openone{\leavevmode\hbox{\small1\kern-3.8pt\normalsize1}}%
\newcommand{\oo}{\openone}
\usepackage{amsfonts}
\newcommand{\ab}[1]{{{\ensuremath{\mathbb{#1}}}}}
\begin{document}
\title{Generalized Taub-NUT metrics and Killing-Yano tensors}
\author{Mihai Visinescu 
\thanks{E-mail:~~~ mvisin@theor1.theory.nipne.ro}\\
{\small \it Department of Theoretical Physics,}\\
{\small \it National Institute for Physics and Nuclear Engineering,}\\
{\small \it P.O.Box M.G.-6, Magurele, Bucharest, Romania}}
%\date{\today}
\date{ }
\maketitle

\begin{abstract}
A necessary condition that a St\"ackel-Killing tensor of valence 2 
be the contracted product of a Killing-Yano tensor of valence 2 with 
itself is re-derived for a Riemannian manifold. This condition is applied 
to the generalized Euclidean Taub-NUT metrics which admit a Kepler type 
symmetry. It is shown that in general the St\"ackel-Killing tensors
involved in the Runge-Lenz vector cannot be expressed as a product of 
Killing-Yano tensors. The only exception is the original Taub-NUT metric.

~

Pacs 04.20.Me

\end{abstract}

\section{Introduction}
The Euclidean Taub-NUT metric is involved in many modern studies in 
phy\-sics. 
Hawking \cite{Ha} has suggested that the Euclidean Taub-NUT metric 
might give rise to the gravitational analog of the Yang-Mills 
instanton. In this case Einstein's equations are satisfied with 
zero cosmological constant and the manifold is {\ab R}$^4$ with a 
boundary which is a twisted three-sphere $S^3$ possessing a distorted 
metric.
The Kaluza-Klein monopole was obtained by embedding the Taub-NUT 
gravitational instanton into five-dimensional Kaluza-Klein theory. On the 
other hand, in the long-distance limit, neglecting radiation, the 
relative motion of two monopoles is described by the geodesics of this 
space \cite{Ma,AH}. 

From the symmetry viewpoint, the geodesic motion in Taub-NUT space 
admits a ``hidden" symmetry of the Kepler type  if a cyclic variable 
is gotten rid of \cite{GM,GR,FH,CFH}. In general the ``hidden" symmetries 
of the manifold manifest themselves as St\" ackel-Killing tensors of 
valence $r>1$ \cite{GH}. The conserved quantities along geodesics are 
homogeneous functions in momentum $p_\mu$ of degree $r$, and which 
commute with the Hamiltonian 
\be\label{H}
H = {1\over 2} g^{\mu\nu} p_\mu p_\nu
\ee
in the sense of Poisson brackets.

In the Taub-NUT geometry there are four Killing-Yano tensors \cite{Ya}. 
Three of these are complex structure realizing the quaternionic algebra 
and the Taub-NUT manifold is hyper-K\" ahler \cite{GR}. In addition to 
these three vector-like Killing-Yano tensors, there is a scalar one which 
has a non-vanishing field strength and it exists by virtue of the 
metric being type $D$.

For the geodesic motions in the Taub-NUT space, the conserved vector 
analogous to the Runge-Lenz vector of the Kepler type problem  is 
quadratic in 4-velocities, its components are 
St\" ackel-Killing tensors and they can be expressed as symmetrized 
products of  Killing-Yano tensors \cite{GR,vH1,VV1,VV2}.

The Killing-Yano tensors play an important role in 
the models for relativistic spin one half particles involving 
anti-commuting vectorial degrees of freedom, usually called the 
spinning particles \cite{BM,BCL,GRH,vH2}. The configuration space of 
spinning particles (spinning space) is an extension of an ordinary 
Riemannian manifold, parametrized by local 
coordinates {$\{$}$x^\mu${$\}$}, to a graded manifold parametrized by 
local coordinates {$\{$}$x^\mu, \psi^\mu${$\}$}, with the first set of 
variables being Grassmann-even (commuting) and the second set 
Grassmann-odd (anti-commuting). In the spinning case the generalized 
Killing equations are more involved and new procedures have been 
conceived \cite{GRH,VV2}. In particular, if the Killing tensors can be 
written in terms of Killing-Yano tensors ( and that is the case of the 
Taub-NUT space), the generalized Killing equations can be solved 
explicitly in a simple, closed form. 

In the last time, Iwai and Katayama \cite{IK1,IK2,IK3,YM} extended 
the Taub-NUT metric so that it still admits a Kepler-type symmetry. 
This class of metrics, of course, includes the original Taub-NUT 
metric. 

The aim of this paper is to investigate if the St\" ackel-Killing tensors 
involved in the conserved Runge-Lenz vector of the extended Taub-NUT 
me\-trics can also be expressed in terms of Killing-Yano tensors.

The relationship between Killing tensors and Killing-Yano tensors has 
been investigated to the purpose of the Lorentzian geometry used in 
general relativity \cite{Co,DR}. In the next section we re-examine 
the conditions that a Killing tensor of valence $2$ be the contracted 
product of a Killing-Yano tensor of valence $2$ with itself. The 
procedure is quite simple and devoted to the Riemannian geometry 
appropriate to Euclidean Taub-NUT metrics.

In Section 3 we show that in general the Killing tensors involved in 
the Runge-Lenz vector cannot be expressed as a product of Killing-Yano 
tensors. The only exception is the original Taub-NUT metric.

Our comments and concluding remarks are presented in Section 4.

\section{The relationship between Killing tensors and Killing-Yano tensors}
We consider a $4-$dimensional Riemannian manifold $M$ and a    
metric $g_{\mu\nu}(x)$ on $M$ in local coordinates $x^\mu$. We write 
the metric in terms of the local orthonormal vierbein frame 
$e^a_{~\mu}$
\be\label{ds}
ds^2 = g_{\mu\nu}(x) dx^\mu dx^\nu = \sum_{a=0,1,2,3} (e^a)^2
\ee
where $e^a = e^a_{~\mu} dx^\mu$. Greek indices $\mu, \nu, ...$ are 
raised and lowered with $g_{\mu\nu}$ or its inverse $g^{\mu\nu}$, while 
Latin indices $a, b, ...$ are raised and lowered by the flat metric 
$\delta_{ab}, a,b = 0,1,2,3$. Vierbeins and inverse vierbeins 
inter-convert Latin and Greek indices when necessary.

The following two generalization of the Killing vector equation have 
become of interest in physics \cite{DR}:\\
(a) A  tensor $f_{\mu_1...\mu_r}$ 
is called a Killing-Yano tensor of valence $r$ if it is totally 
anti-symmetric and it satisfies the equation
\be\label{yano}
f_{\mu_1...(\mu_r;\lambda)} = 0.
\ee
(b) A symmetric tensor field $K_{\mu_1...\mu_r}$  is called a 
St\" ackel-Killing tensor of valence $r$ iff
\be\label{sk}
K_{(\mu_1...\mu_r;\lambda)} = 0.
\ee

Let $\Lambda ^2$  be the space of two-forms $\Lambda ^2 := \Lambda ^2 
T^* ({\ab R}^4 - \{ 0 \})$. We  define self-dual and anti-self dual 
bases for $\Lambda ^2$ using the vierbein one-forms $e^a$ \cite{EGH}:
\be\label{basis}
 basis~ of~~~ \Lambda^{2}_{\pm} = \left\{ \begin{array}{lr}
\lambda^1_{\pm} = e^0 \wedge e^1 \pm  e^2 \wedge e^3& \\
\lambda^2_{\pm} = e^0 \wedge e^2 \pm  e^3 \wedge e^1~~, ~~~~& 
*\lambda^i_{\pm} 
= \pm \lambda^i_{\pm} \\
\lambda^3_{\pm} = e^0 \wedge e^3 \pm  e^1 \wedge e^2& 
\end{array}
\right. 
\ee

Let $f$ be a Killing-Yano tensor of valence 2 and  $*f$ its dual. 
The symmetric combination of $f$ and $*f$ is a self-dual two-form
\be
f + *f = \sum_{i=1,2,3} y_i \lambda^i_+
\ee
while their difference is an anti-self-dual two-form
\be
f - *f = \sum_{i=1,2,3} z_i \lambda^i_-.
\ee

An explicit evaluation shows that 
\be
(f + *f)^2  = - \sum_{i=1,2,3} (y_i)^2 \cdot \oo,
\ee
\be
(f - *f)^2  = - \sum_{i=1,2,3} (z_i)^2 \cdot \oo
\ee
where \oo ~is $4\times$4 identity matrix.

Let us suppose that a St\" ackel-Killing tensor $K_{\mu\nu}$ can be 
written as the contracted product of a Killing-Yano tensor 
$f_{\mu\nu}$ with itself:
\be\label{kyy}
K_{\mu\nu} = f_{\mu\lambda} \cdot f^{\lambda}_{~~\nu} = 
(f^2)_{\mu\nu}~,~~ \mu , \nu = 0,1,2,3.
\ee
We infer from the last equations that: 
\be
K + {1\over 16} \left[\sum_{i} (y^2_i - z^2_i)\right]^2 K^{-1} + 
{1\over 2} \sum_{i} (y^2_i + z^2_i) \cdot  \oo = 0.
\ee

On the other hand the Killing tensor $K$ is symmetric and it can be 
diagonalized with the aid of an orthogonal matrix. Its eigenvalues 
satisfy an equation of the second degree:
\be
\lambda^{~2}_\alpha + {1\over 2} \sum_{i} (y^2_i + z^2_i) 
\lambda_\alpha + {1\over 16}\left[ \sum_{i} (y^2_i - z^2_i)\right]^2 
= 0
\ee
with at most two distinct roots.

In conclusion a St\" ackel-Killing tensor $K$ which can be written as 
the square of a Killing-Yano tensor has at the most two distinct 
eigenvalues.

\section{Generalized Taub-NUT metrics}
For a special choice of coordinates the generalized Euclidean Taub-NUT 
metric considered by Iwai and Katayama \cite{IK1,IK2,IK3,YM} takes the 
form:
\be\label{dg}
ds^2_G=f(r)[dr^2+r^2d\theta^2+r^2\sin^2\theta\, d\vf^2]
+g(r)[d\chi+\cos\theta\, d\vf]^2 
\ee
where $r>0$ is the radial coordinate of $ \ab{R}^4 - \{0\}$ ,
the angle variables $(\theta,\vf,\chi), (0\leq\theta<\pi, 0\leq\vf<2\pi, 
0\leq\chi<4\pi)$ parameterize the unit sphere $S^3$, and $f(r)$ and 
$g(r)$ are arbitrary functions of $r$.

We decompose the metric (\ref{dg}) into the orthogonal vierbein basis:
\ba\label{ea}
e^0 &=& g(r)^{1\over 2} ( d\chi + \cos\theta d\vf) ,\no
e^1 &=& r f(r)^{1\over 2} (\sin\chi d\theta - \sin\theta \cos\chi 
d\vf) ,\no
e^2 &=& r f(r)^{1\over 2} (-\cos\chi d\theta - \sin\theta \sin\chi 
d\vf) ,\no
e^3 &=& f(r)^{1\over 2} dr.
\ea

Spaces with a metric of the form above have an isometry group 
$ SU(2)\times U(1)$. The four Killing vectors are
\be
D_A=R_A^\mu\,\partial_\mu,~~~~A=0,1,2,3,
\ee
where
\ba
D_0&=&{\partial\over\partial\chi},\no
D_1&=&-\sin\vf\,{\partial\over \partial\theta}-\cos\vf\,\cot\theta
\,{\partial\over\partial\vf}+{\cos\vf\over\sin\theta}\,{\partial\over
\partial\chi},\no
D_2&=&\cos\vf\,{\partial\over \partial\theta}-\sin\vf\,\cot\theta
\,{\partial\over\partial\vf}+{\sin\vf\over\sin\theta}\,{\partial\over
\partial\chi},\no
D_3&=&{\partial\over\partial\vf}.
\ea

$D_0$ which generates the $U(1)$ of $\chi$ translations, commutes 
with the other Killing vectors. In turn the remaining three vectors,
corresponding to the invariance of the metric (\ref{dg}) under spatial
rotations ($A=1,2,3$), obey an $SU(2)$ algebra with
\be
[D_1, D_2]=-D_3~,~{\it etc... }.
\ee

Let us consider geodesic flows of the generalized Taub-NUT metric which 
has the Lagrangian $L$ on the tangent bundle $T({\ab R}^4 - \{0\})$
\be \label{lag}
L = {1\over 2} f(r) [\dot r^2 + r^2(\dot \theta^2 + \sin^2\theta 
\dot\vf^2)] + {1\over 2} g(r) (\dot\chi + \cos\theta \dot\vf)^2 
\ee
where$(\dot r,\dot\theta,\dot\vf,\dot\chi,r,\theta,\vf,\chi)$ stand for 
coordinates in the tangent bundle. Since $\chi$ is a cyclic variable
\be \label{q}
q = g(r) (\dot\theta + \cos\theta \dot\vf)
\ee
is a conserved quantity. This is known in the literature as the 
``relative electric charge".

Taking into account this cyclic variable, the dynamical system for the 
geodesic flow on $T({\ab R}^4 - \{ 0 \})$ can be reduced to a system on
$T({\ab R}^3 - \{ 0 \})$. The reduced system admits manifest rotational 
invariance, and hence has a conserved angular momentum:
\be
\vect{J}=\vect{r}\times\vect{p}\,+\,q\,{\vect{r}\over r} 
\ee
where $\vect{r}$ denotes the three-vector $\vect{r} = (r,\theta,\vf)$ and 
$\vect{p} = f(r)\dot{\vect{r}}$ is the me\-cha\-ni\-cal momentum.

If $f(r)$ and $g(r)$  are taken to be
\be \label{fg}
f(r) = {4 m + r\over r} ~~~, ~~~g(r) = {16 m^2 r\over 4 m + r}
\ee
the metric $ds^2_G$ becomes the original Euclidean 
Taub-NUT metric. 
As observed in \cite{GR}, the Taub-NUT geometry also possesses four 
Killing-Yano tensors of valence 2. The first three are rather special: 
they are covariantly constant (with vanishing field strength)
\ba\label{fi}
f_i &=&8m(d\chi + \cos\theta d\varphi)\wedge dx_i -
\epsilon_{ijk}(1+\frac{4m}{r}) dx_j \wedge dx_k,\no
D_\mu f^\nu_{i\lambda} &=&0~, ~~~~i,j,k=1,2,3.
\ea

They are mutually anti-commuting and square the minus unity:
\be
f_i f_j + f_j f_i = -2\delta_{ij}.
\ee

Thus they are complex structures realizing the quaternion algebra. 
Indeed, the Taub-NUT manifold defined by (\ref{dg}) and (\ref{fg}) 
is hyper-K\" ahler. 

In addition to the above vector-like Killing-Yano tensors there also is 
a scalar one
\be\label{fy}
f_Y =8m(d\chi + \cos\theta  d\varphi)\wedge dr +
4r(r+2m)(1+\frac{r}{4m})\sin\theta  d\theta \wedge d\varphi
\ee
which has a non-vanishing component of the field strength 
\be
{f_{Y}}_{r\theta;\varphi} = 2(1+\frac{r}{4m})r\sin\theta.
\ee

In the original Taub-NUT case there is a conserved vector analogous 
to the Runge-Lenz vector of the Kepler-type problem:
\be\label{knut}
\vect{K} = \frac{1}{2} \vect{K}_{\mu\nu}\dot x^\mu\dot x^\nu = 
\vect{p}\times\vect{j} + \left(\frac{q^2}{4m}-
4mE\right)\frac{\vect r}{r}
\ee
where the conserved energy $E$, from eq. (\ref{H}), is 
\be\label{E}
E = {\vect{p}^{~2}\over 2 f(r)} + {q^2 \over 2 g(r)} .
\ee

The components $K_{i\mu\nu}$ involved with the Runge-Lenz type vector 
(\ref{knut}) are Killing tensors and they can be expressed as 
symmetrized products of the Killing-Yano tensors $f_i$ (\ref{fi}) and
$f_Y$ (\ref{fy}) \cite{VV1,VV2}: 
\be\label{kten}
K_{i\mu\nu} - \frac{1}{8m} (R_{0\mu} R_{i\nu} + R_{0\nu} R_{i\mu}) = 
m\left( f_{Y\mu\lambda}
{{f_{i}}^\lambda}_\nu + f_{Y\nu\lambda} {{f_{i}}^\lambda}_\mu
\right). 
\ee

Returning to the generalized Taub-NUT metric, on the analogy of 
eq.(\ref{knut}), Iwai and Katayama \cite{IK1,IK2,IK3,YM} assumed that 
in addition to the angular momentum vector there exist a conserved 
vector $\vect{S}$ of the following form: 
\be\label{kik}
\vect{S} = \vect{p} \times \vect{J} + \kappa {\vect{r} \over r}
\ee
with an unknown constant $\kappa$.

It was found that the metric (\ref{dg}) still admits a Kepler type 
symmetry (\ref{kik}) if the functions $f(r)$ and $g(r)$ take, 
respectively, the form
\be\label{fgg}
f(r) = { a + b r\over r}  ~~,~~ g(r) = { a r + b r^2 \over 
1 + c r + d r^2}
\ee
where $a,b,c,d $ are constants. The constant $\kappa$ involved in the 
Runge-Lenz vector (\ref{kik}) is
\be
\kappa = - a\,E + {1\over 2} c\,q^2.
\ee

If $ab >0$ and $c^2 -4 d < 0$ or $ c > 0 , d > 0$, no singularity of 
the metric appears in $\ab{R}^4 - \{0\}$. On the other hand, if $ab < 
0$ a manifest singularity appears at $ r = - a/b$ \cite{IK2}.

It is straightforward to verify that the components of 
the vector $\vect{S}$  are St\" ackel-Killing tensors in 
the extended Taub-NUT space (\ref{dg}) with the function $f(r)$ and 
$g(r)$  given by (\ref{fgg}).
Moreover the Poisson brackets between the components of $\vect{J}$ and 
$\vect{S}$ are \cite{IK1}:
\ba\label{jj} 
\{J_i, J_j\} &=& \epsilon_{ijk} J_k ,\no
\{J_i, S_j\} &=& \epsilon_{ijk} S_k ,\no
\{S_i, S_j\} &=& (d\,q^2 - 2\,b\,E)\epsilon_{ijk} J_k
\ea
as it is expected from the same relations known for the original 
Taub-NUT metric.

Our task is to investigate if the components of the Runge-Lenz vector 
(\ref{kik}) can be the contracted product of Killing-Yano tensors of 
valence $2$. On the model of eq.(\ref{kten})  from the original 
Taub-NUT case it is not required that a component $S_i$ of the 
Runge-Lenz vector (\ref{kik}) to be directly expressed  as a 
symmetrized product of Killing-Yano tensors. Taking into account that 
$\vect{S}$ transforms as a vector under rotations generated by 
$\vect{J}$, eq.(\ref{jj}), the components $S_{i\mu\nu}$ can be 
combined with trivial St\" ackel-Killing tensors of the form
$ (R_{0\mu} R_{i\nu} + R_{0\nu} R_{i\mu})$ to get the appropriate 
tensor which has to be decomposed in a product of Killing-Yano tensors.

In order to use the results from the previous  section, we shall write 
the symmetrized product of two different Killing-Yano tensors $f'$ and 
$f''$ as a contracted product of $ f' + f''$ with itself, 
extracting adequately the contribution of $f'^2$ and $f''^2$. 
Since the generalized Taub-NUT space (\ref{dg}) does not admit any 
other non-trivial St\" ackel-Killing tensor besides the metric 
$g_{\mu\nu}$ and the components $S_{i\mu\nu}$ of (\ref{kik}),
$f'^2 $  and $f''^2$  should be connected with the scalar conserved 
quantities $E, \vect{J}^2, q^2$ through the tensors $g_{\mu\nu},
\sum_{A=1,2,3} R_{A\mu}R_{A\nu}$ and $R_{0\mu}R_{0\nu}$.

In conclusion we shall consider a general linear combination between a 
component $S_i$ of the Runge-Lenz vector (\ref{kik}) and symmetrized 
pairs of Killing vectors of the form
\be\label{comb}
S_{iab} + \alpha_1 \sum_{A=1}^{3} R_{Aa}R_{Ab} + 
\alpha_2 R_{0a}R_{0b} + \alpha_3 (R_{0a}R_{ib} + R_{ia}R_{0b})
\ee
where $\alpha_{i}$ are constants. We are looking for the conditions the 
above tensor be the contracted product of a Killing-Yano tensor with 
itself. For this purpose we evaluate the eigenvalues of the matrix 
(\ref{comb}) and we get that it has at the most two distinct 
eigenvalues if and only if 
\ba\label{cd}
\alpha_1 + \alpha_2 = 0 ,\no
\alpha_3 = - {c\over 4} ,\no
d = {c^2\over 4}.
\ea

For example, if the above conditions are satisfied, the eigenvalues of 
the matrix (\ref{comb}) for the third component $S_3$ of the Runge-Lenz 
vector (\ref{kik}) are
\be\label{vap1}
\lambda_1 = {1\over 2}\left( b r \cos\theta + ( a + b r) \left(r 
\alpha_1  + \sqrt{ 1 + r^2 \alpha_1^2 + 2 r \alpha_1 \cos\theta}
\right)\right)
\ee
with the eigenvectors
\ba\label{vep1}
\left\{ \tan\chi, \left(r \alpha_1 + \cos\theta 
+ \sqrt{ 1 + r^2 \alpha_1^2 + 2 r \alpha_1 \cos\theta}\right)
\csc\theta \sec\chi, 0, 1\right\} ,\no
\left\{\left( (r \alpha_1 + \cos\theta 
- \sqrt{ 1 + r^2 \alpha_1^2 + 2 r \alpha_1 \cos\theta}\right)
\csc\theta \sec\chi, - \tan\chi, 1, 0\right\}
\ea
and
\be\label{vap2}
\lambda_2 = {1\over 2}\left( b r \cos\theta + ( a + b r) \left(r 
\alpha_1  - \sqrt{ 1 + r^2 \alpha_1^2 + 2 r \alpha_1 \cos\theta}
\right)\right)
\ee
with the eigenvectors
\ba\label{vep2}
\left\{ \tan\chi, \left(r \alpha_1 + \cos\theta 
- \sqrt{ 1 + r^2 \alpha_1^2 + 2 r \alpha_1 \cos\theta}\right)
\csc\theta \sec\chi, 0, 1\right\} ,\no
\left\{\left( (r \alpha_1 + \cos\theta 
+ \sqrt{ 1 + r^2 \alpha_1^2 + 2 r \alpha_1 \cos\theta}\right)
\csc\theta \sec\chi, - \tan\chi, 1, 0\right\}.
\ea

Hence the constants involved in the functions $f,g$  are 
constrained, restricting  accordingly their expressions. It is worth to 
mention that if  relation (\ref{cd}) between the constants $c$ and 
$d$ is satisfied, the metric is conformally self-dual or 
anti-self-dual depending upon the sign of the quantity $2 + cr$ 
\cite{IK2}. More precisely, for the Weyl curvature tensor
\be \label{weyl}
C^i_{jkl} = R^i_{jkl} - {1\over 2}( \delta^i_k R_{jl}-\delta^i_l R_{jk}
+ \delta^j_l R_{ik} - \delta^j_k R_{il} ) + {1\over 6} R (\delta^i_k 
\delta_{jl} - \delta^i_l \delta_{jk})
\ee
one can define a two-form
\be \label{wij}
W_{ij} = {1\over 2} \sum_{k,l} C^i_{jkl} e^k\wedge e^l.
\ee

With respect to basis (\ref{basis}) the representation matrix $W$ of 
(\ref{wij}) takes the block diagonal form
\be
W = \left( \begin{array}{cc}
W^+& 0\\
0&W^-
\end{array} \right)
\ee
where $W^+$ and $W^-$ are $3\times 3$ matrices representing the induced 
linear transformation of the invariant subspaces $\Lambda^2_+$ and 
$\Lambda^2_-$ respectively..
If the constants $c$ and $d$ satisfy (\ref{cd}), the extended Taub-NUT 
metric (\ref{fgg}) with $2 + c r > 0$ is conformally self-dual and one 
has \cite{IK2}
\be
W^+ = { c \over 2( a + b r) ( 1 + c r /2)^2} W_0 ~~,~~~ W^- = 0
\ee
where $W_0$ is a diagonal matrix
\be
W_0 =\left( \begin{array}{ccc}
-1& &\\
& -1&\\
&&2
\end{array} \right).
\ee 
For $2 + c r < 0$, the metric is conformally anti-self-dual and the 
expressions of $W^+$ and $W^-$ are interchanged.

Finally the condition stated for a St\" ackel-Killing tensor to be 
written as the square of a skew symmetric tensor in the form (\ref{kyy}) 
must be supplemented with  eq.(\ref{yano}) which defines a 
Killing-Yano tensor. To verify this last 
condition we shall use the Newman-Penrose formalism for Euclidean 
signature \cite{AN}. We introduce a tetrad which will be given as an 
isotropic complex dyad defined by the vectors $l,m$ together with their 
complex conjugates subject to the normalization conditions
\be\label{tet}
l_\mu \bar{l}^\mu = 1, ~~~m_\mu \bar{m}^\mu = 1
\ee
with all others vanishing and the metric is expressed in the form
\be
ds^2 = l\otimes \bar{l} + \bar{l}\otimes l + m\otimes \bar{m} + 
\bar{m}\otimes m.
\ee
 
For a St\" ackel-Killing tensor $K$ with two distinct eigenvalues one 
can choose the tetrad in such that
\be
K_{\mu\nu} = 2 \lambda_1^2 l_{(\mu}\bar{l}_{\nu)} + 
2 \lambda_2^2 m_{(\mu}\bar{m}_{\nu)}.
\ee
The skew symmetric tensor $f_{\mu\nu}$ which enter decomposition
(\ref{kyy}) has the form
\be\label{ky}
f_{\mu\nu} = 2 \lambda_1 l_{[\mu}\bar{l}_{\nu]} + 
2 \lambda_2 m_{[\mu}\bar{m}_{\nu]}. 
\ee

Taking again the example of the third component $S_3$, the eigenvalues 
$\lambda_1$ and $\lambda_2$ are given by (\ref{vap1}) and (\ref{vap2}) 
and  the tetrad (\ref{tet}) can be inferred from the eigenvectors 
(\ref{vep1}) and (\ref{vep2}) through a standard orthonormalization 
procedure. Finally, imposing eq.(\ref{yano}), we get that (\ref{ky}) 
is a Killing-Yano tensor only if
\be\label{abc}
c = {2 b\over a}.
\ee

With this constraint, together with (\ref{cd}), the extended metric 
(\ref{dg}) coincides, up to a constant factor, with the original 
Taub-NUT metric on setting $ a/b = 4 m$. Note that in 
eqs.(\ref{vap1})-(\ref{vep2}) the constant $\alpha_1$ is not fixed. In 
fact, the product of two Killing-Yano tensors $f'\cdot f''$ is 
invariant under the rescaling $f'\rightarrow \alpha f',
f''\rightarrow {1\over \alpha} f''$. Choosing adequately the 
normalization of the Killing-Yano tensors, for $ \alpha_1 = -{1\over 
4m}$ we recover precisely the original Taub-NUT decomposition 
(\ref{kten}) with $f' = f_i$ and $f'' = f_Y$ normalized as in (\ref{fi}) 
and (\ref{fy}).

\section{Concluding remarks}
The aim of this paper is to show that the extensions of the Taub-NUT 
geometry do not admit a Killing-Yano tensor, even if they possess 
St\" ackel-Killing tensors.

This result is not unexpected. The conserved quantities $K_{i\mu\nu}$
which enter eq.(\ref{kten}) are the components of the Runge-Lenz vector 
$\vect{K}$ given in (\ref{knut}). In the original Taub-NUT case these 
components $K_{i\mu\nu}$ are related to the symmetrized products between 
the Killing-Yano tensors $f_i$ (\ref{fi}) and $f_Y$ (\ref{fy}). 
Adequately the three Killing-Yano tensors $f_i$ transform as vectors 
under rotations generated by $\vect{J}$ like the Runge-Lenz vector 
(\ref{jj}), while $f_Y$ is a scalar.

The extended Taub-NUT metrics are not Ricci flat and, consequently,
not hyper-K\" ahler. On the other hand the existence of the 
Killing-Yano tensors $f_i$ is correlated with the hyper-K\" ahler, 
self-dual structure of the metric.

The in-existence of the Killing-Yano tensors makes more laborious the 
study of ''hidden" symmetries in
models of relativistic particles with spin involving anti-commuting 
vectorial degrees of freedom. In general the conserved quantities from 
the scalar case receive a spin contribution involving an even number of 
Grassmann variables $\psi^\mu$. For example, starting with a Killing 
vector $K_\mu$, the conserved quantity in the spinning case is
\be
J(x,\dot x,\psi) = K^\mu \dot x_\mu + {i\over 2} K_{[\mu;\nu]}
\psi^\mu \psi^\nu.
\ee
The first term in the r.h.s. is the conserved quantity in the scalar
case, while the last term represents the contribution of the spin.

A ``hidden" symmetry is encapsulated in a St\" ackel-Killing tensor of 
valence $r>1$. The generalized Killing equations on spinning spaces 
including a St\" ackel-Killing tensor are more involved. Unfortunately 
it is not possible to write closed, analytic expressions of the 
solutions of these equations using directly the components of the 
St\" ackel-Killing tensors. However, assuming that 
the St\" ackel-Killing tensors can be written as symmetrized products 
of pairs of Killing-Yano tensors, the evaluation of the spin corrections 
is feasible \cite{GRH,VV1,VV2,vH2}.
 
If the Killing-Yano tensors are missing, to take up the 
question of the existence of extra supersymmetries and the relation 
with the constants of motion we are forced to enlarge the approach to 
Killing equations (\ref{yano}), (\ref{sk}). In fact, in ref.\cite{GRH}, 
supersymmetries are shown to depend on the existence of a  tensor field  
$f_{\mu\nu}$ satisfying eq.(\ref{yano}) which will be referred to as the 
$f$-symbol. The general conditions for constants of motion were derived, 
and it was shown that one can have new supercharges which do not commute 
with the original supercharge $Q = \dot x_\mu \psi^\mu$ if one allows the 
$f$-symbols to have a symmetric part. It was shown that in this case 
the anti-symmetric part does not satisfy the Killing-Yano condition 
(\ref{yano}). We should like to remark that the general conditions of 
ref.\cite{GRH} allow more possibilities than Killing-Yano tensors for 
the construction of supercharges.

Summing up, we believe that the relation between the $f$-symbols and  
the Killing-Yano tensors could be fruitful and that it should deserve 
further studies. An analysis of the $f$-symbols in the generalized 
Taub-NUT geometry is under way \cite{MV}.

\subsection*{Acknowledgments}
I would like to thank Luca Lusanna for interesting and helpful 
conversations on the present topic and Diana Vaman for discussions at 
an early stage of this work. 
This investigation has been supported in part by the Rumanian Academy 
grants  GR-68/1999, GR-01/2000.

\end{document}